\documentclass[usedcolumn]{raa}

\usepackage{graphicx,times}             
\usepackage{amssymb,amsmath}
\usepackage[a4paper=true,dvipdfm=true,pagebackref=true]{hyperref}
\hypersetup{colorlinks = true, linkcolor = green, anchorcolor = red, citecolor = blue, filecolor = red, pagecolor = red, urlcolor = red}

\begin{document}

   \title{The redshift distribution of BL Lacs and FSRQs}

   \volnopage{Vol.0 (20xx) No.0, 000--000}      %%preserved for Editor. DOn't remove!
   \setcounter{page}{1}          %%starting page, preserved for Editor. DOn't remove!

   \author{David Garofalo
      \inst{1}
   \and Chandra B. Singh
      \inst{2}
   \and Dylan T. Walsh
      \inst{3}
   \and Damian J. Christian
     \inst{3}
   \and Andrew M. Jones
      \inst{4}
   \and Alexa Zack
      \inst{1}
   \and Brandt Webster
      \inst{1}
   \and Matthew I. Kim
      \inst{3}
   }

   \institute{Department of Physics, Kennesaw State University, Marietta GA 30060, USA; {\it dgarofal@kennesaw.edu.}\\
        \and
            The Raymond and Beverly Sackler School of Physics and Astronomy, Tel Aviv University, Tel Aviv 69978, Israel\\
        \and
            Department of Physics \& Astronomy, California State University, Northridge CA 91330, USA \\
        \and            
            Boise State University, Computational Science \& Engineering, Boise ID 83725, USA \\
\vs\no
   {\small Received~~20xx month day; accepted~~20xx~~month day}}

\abstract{ Flat spectrum radio quasars (FSRQs) and BL Lacs are powerful jet producing active galactic nuclei associated with supermassive black holes accreting at high 
and low Eddington rates, respectively. Based on the Millennium Simulation, Gardner \& Done (2014; 2018) have predicted their redshift distribution by appealing to ideas 
from the spin paradigm in a way that exposes a need for a deeper discussion on three interrelated issues: (1) an overprediction of BL Lacs compared to flat spectrum 
radio quasars; (2) a difference in FSRQ and BL Lac distributions; (3) a need for powerful but different jets at separated cosmic times. Beginning with Gardner \& Done's 
determination of Fermi observable FSRQs based on the distribution of thermal accretion across cosmic time from the Millennium Simulation, we connect FSRQs to BL Lacs by 
way of the gap paradigm for black hole accretion and jet formation to address the above issues in a unified way. We identify a physical constraint in the paradigm for 
the numbers of BL Lacs that naturally leads to separate peaks in time for different albeit powerful jets. In addition, we both identify as puzzling and ascribe physical 
significance to, a tail end in the BL Lac curve versus redshift that is unseen in the redshift distribution for FSRQs.
}

   \authorrunning{David Garofalo et al.}
   \titlerunning{The redshift distribution of BL Lacs and FSRQs}  

   \maketitle

\section{Introduction}         
\label{sect:intro}
FSRQs are a subclass of the blazar group of powerful jet emitting active galactic nuclei with strong emission lines (Urry \& Padovani 1995) and jet morphologies belonging to the 
FRII class (Fanaroff \& Riley 1974) that peak around redshift 1-2 and drop off below and above that (Ajello et al 2012; Mao et al 2017). FSRQs are near-Eddington accretors and 
likely standard  radiatively efficient disks. The optical emission lines are very prominent and appear in both narrow (Narrow Line Radio Galaxies) and broad (Broad Line Radio 
Galaxies) form that are the parent family of FSRQs. The parent family of BL Lacs is thought to be the FRI radio galaxy group. BL Lacs are a subclass of the blazar group of 
powerful jet emitting active galactic nuclei with weak or no emission lines (Stickel et al 1991; Urry \& Padovani 1995) and jet morphologies belonging to the
FRI class (Fanaroff \& Riley 1974) that peak at lower redshift. They are highly sub-Eddington accretors and likely radiatively inefficient advection dominated disks. 
The parent families of both groups have strong radio/optical/soft X-ray correlations with optical emission being jet related. Whereas the FSRQs display a strong time evolution, 
the BL Lacs show little to no change over time.\\

Gardner \& Done (2014; 2018; henceforth GD14 and GD18) identify from the Fermi Large Area Telescope 300 FSRQs and 500 BL Lacs (Abdo et al. 2010; Ackermann et al. 2011) whose 
distributions as a function of redshift they attempt to reproduce from theory. By appealing to the Millennium Simulation (Springel et al. 2005, Fanidakis et al. 2011; 2012 - Figure 2)
, GD14 and GD18 identify the range of accretion across the black hole mass scale to determine as a function of redshift the number of objects that accrete above and below the critical
accretion rate in terms of the Eddington accretion rate at 0.01 $(dM/dt)_{Edd}$. This constitutes the theoretical boundary between radiatively efficient thermal accretion 
(Shakura \& Sunyaev 1973) and advection dominated accretion (Narayan \& Yi 1995). By evaluating the subset of all accreting black holes whose characteristics allows them to be Fermi 
detectable FSRQs and BL Lacs, they begin with the assumption that FSRQs and BL Lacs differ solely by the accretion rate such that above the critical limit all objects are FSRQs while
below it they are BL Lacs. They find the predicted number of FSRQs and BL Lacs to be two and three orders of magnitude larger than the observed number, respectively. In other words, 
the subset of radiatively inefficient accretion dominates the radiatively efficient ones by one order of magnitude. In order to improve the match with observations, GD14 and GD18 
further constrain this subset of accreting black holes by the dimensionless black hole spin such that for a $>$ 0.8 and dM/dt $<$ $0.01(dM/dt)_{Edd}$ , all objects are BL Lacs while 
for a $>$ 0.77 and dM/dt $>$ 0.01$(dM/dt)_{Edd}$, all objects are FSRQs. This produces a better fit with observations.\\

The framework presented by GD14 and GD18 raises a series of interconnected issues. First, there is a long history of both theoretical and numerical work suggesting that high 
or even very high black hole spin is needed to explain the most powerful jets in AGNs (e.g. Tchekhovskoy et al. 2010). Hence, there is tension between theory and the 
prescriptions for the jets of most of the BL Lacs as well as most of the FSRQs in both GD14 and GD18. Even if one relaxes the assumption of $a \sim 1$ from theory and simulation 
and allows powerful jets for a $\geq$ 0.8, the conditions that are assumed to produce such spins appear contrived, with mergers and subsequent chaotic accretion when and where 
low spins are necessary such as for the weak and jetless AGN population but with random injection of mergers also contributing to high spins in order to explain the appearance 
of AGN with jets. In fact, signatures of mergers assumed to give the BL Lacs their high spins are not observed. This problem of the separate peaks in the distributions of powerful
jet-producing AGNs (FSRQ at higher redshift compared to BL Lacs) is a longstanding issue for models of jets based on black hole spin. They appear to force one to consider random
pockets of high black hole spin. And, finally, the observations show a tail in the distribution of the BL Lacs (Shaw et al 2013) that is not explained in GD14 - or anywhere else 
for that matter - that we think may be of physical significance.\\

In this paper we attempt the same project of GD14 and GD18 but with an appeal to the gap paradigm for black hole accretion and jet formation (Garofalo, Evans \& Sambruna 2010)
which allows for the possibility of an evolution in black hole spin that is high at some moment in time (but retrograde) and at later times becomes high again (but prograde) 
as a result of the physical mechanism of prolonged accretion that spins black holes down while in retrograde configurations but spins them up in prograde ones. Because we model
FSRQs as retrograde accreting black holes and BL Lacs as prograde accreting black holes, there is a model-dependent natural connection between the two families of objects and an 
expected redshift difference in their peaks. The same connection that explains high spins at different redshift, therefore, also explains why the number of observed FSRQs is of 
the same order as those observed for the BL Lacs, the latter being connected to the former as offspring. Finally, this framework also suggests that the observed tail in the BL 
Lac distribution is a feature of the connection to their progenitor FSRQs. This feature, as far as we know, has not been identified nor explained, but if physical, leads to 
interesting insights and constraints concerning the two AGN families. In Section 2 we model FSRQs, in Section 3 the BL Lacs. In Section 4 we discuss the issues raised 
above and in Section 5 we conclude.

\section{FSRQs}
\label{sect:fsrqs}
Under the assumption that only the subset of black holes accreting at above 0.01 the Eddington rate and appropriate orientation to be detected, are FSRQs (Figure 1), GD18 
overpredict the observed numbers of FSRQs by two orders of magnitude. GD18 further modulate the predicted curve by assuming a restriction on black hole spin, namely that 
FSRQs are produced only for dimensionless black hole spins above 0.77. Their predicted FSRQ number versus redshift is shown in Figure $3$ in green, taken from Figure $5$ in GD18.
In this section we carry out a similar project of determining the predicted number of FSRQs from theory but from the perspective of the gap paradigm (Garofalo, Evans \& Sambruna 2010).
As in GD18 who have taken the results of the Millennium Simulation which provides us with the distribution of accretion rates on black holes of varying mass as a function of redshift, 
our starting point is the black curve in Figure 1. FSRQs are thought to be a subset of the parent family of FRII quasars, which are modeled as radiatively efficient thin disk accretion
around retrograde accreting black holes in the gap paradigm.\\

Prior to considering spin constraints, however, we need the number of black holes accreting above 0.01 Eddington that are in retrograde configuration. Not only do we constrain our 
FSRQ candidates from those black holes accreting above 0.01 Eddington, we further restrict that subset  by selecting black holes whose mass is equal to or greater than 1 billion solar masses. 
A natural assumption might be that post-merger gas funneled toward the black hole will form prograde or retrograde disks randomly so the fraction of retrograde black holes would be random 
and therefore $50\%$ retrograde. This is not correct. Although the details appear elsewhere (Garofalo, Christian \& Jones $2018$ in preparation and references therein) and we limit our 
discussion of this to the results, retrograde black hole configurations appear to be unstable in a way that depends on the mass of the black hole relative to that of the accretion disk and a
body of work suggests how to determine the dependence of that stability on the relative mass (King et al. 2005; Perego et al. 2007; Garofalo et al. 2016). For larger black hole masses, the
range of total mass in the accretion disk will span a wider range and a greater fraction of the total will involve configurations with large black hole mass and small disk mass. 
Although it is expected that even a subset of black holes around $10^{8}$ solar masses may be stable enough to accrete in retrograde mode, the probability of retrograde mode decreases 
among the lower black hole mass population. Because this effect is not well understood, and for simplicity, we assume that only black holes equal to or above $10^{9}$ solar masses 
can accrete in retrograde mode. This will systematically lower our predicted number of FSRQs. However, we will normalize as discussed below. The number of accreting black holes 
equal to or greater than $10^{9}$ solar masses as a function of redshift is given to us by Gardner \& Done and reported in Table 1 column 2. Because the number of radiatively 
efficient accreting black holes clustering around $10^{8}$ solar masses is larger around redshift 1 than redshift 2, our choice of restricting retrograde accretion to $10^{9}$ solar mass
black holes or more, produces an additional systematic effect which is to shift the redshift of the peak of our predicted FSRQ versus redshift plot (Figure 3). As described in detail
later in the discussion, however, our claim is not greater accuracy in our predicted FSRQ and BL Lacs curves as a function of redshift from our phenomenology, rather it is the 
combination of three fundamental physical features  all emerging from the same idea.\\ 

\begin{table}
\begin{center}
\caption[]{Second column is the total number of accreting black holes with $dM/dt > 0.01(dM/dt)_{Edd}$ and mass equal to or above $10^{9}$ solar masses. We assume
half of the objects are in retrograde mode and half of those have sufficient black hole spin values. Hence, to obtain the predicted number of FSRQs you must multiply
the column 2 number by 0.25. We then normalize the maximum value to the observed maximum value – which is 40 – and that requires all numbers be multiplied by 5.56.
In other words, the assumption that only $10^{9}$ solar mass black holes are stable enough for retrograde spin reduces the maximum predicted value by a factor of 5.56.
Fourth column is the observed number of FSRQs. Data provided courtesy of E. Gardner and C. Done.}

\begin{tabular}{c|c|c|c}
  \hline\noalign{\smallskip}
Redshift & Total $M > 10^{9}M_{s}$  & Normalized Predicted No. of FSRQs  & Observed No. of FSRQs \\
  \hline\noalign{\smallskip}
0.0      & 2.34                     & 3.23                               & 20  \\
0.5      & 6.84                     & 9.48                               & 20  \\
1        & 11.34                    & 15.73                              & 40  \\
1.5      & 20.05                    & 27.85                              & 20  \\
1.8      & 28.76                    & 40                                 & 20 \\
2.07     & 17.52                    & 24.35                              & 14 \\
2.5      & 9.26                     & 12.87                              & 10 \\
3.3      & 1                        & 1.39                               & 2 \\
4.18     & 0                        & 0                                  & NA \\
5.28     & 0                        & 0                                  & NA \\
6.7      & 0                        & 0                                  & NA \\
7.88     & 0                        & 0                                  & NA \\
8.55     & 0                        & 0                                  & NA \\
9.27     & 0                        & 0                                  & NA \\
   \noalign{\smallskip}\hline
\end{tabular}
\end{center}
\end{table}

In addition to this, we need to consider black hole spin. Jet power depends on black hole spin as determined in Garofalo, Evans \& Sambruna (2010) Figure 4, where we find the largest 
prograde jet powers at $a \sim 0.9$ are achieved and surpassed in retrograde configurations for $a > 0.3$. Hence, we need to multiply by some fraction for the retrograde accreting 
black hole population because not all of them satisfy these spin requirements to become FSRQs in the model. In other words, we need to identify the numbers of those accreting black 
holes with $a > 0.3$. To accomplish this we incorporate the spin distribution from simulations of mergers (Berti \& Volonteri 2008 and references therein), where it is shown that 
under a representative astrophysical parameter space for merging black holes, the final dimensionless spin magnitude averages near a = 0.7 with maximal spins very unlikely (i.e. $a > 0.9$). 
Hence, we are effectively modeling the FSRQs in a spin range $0.3 < a < 0.9$ and determine from Berti \& Volonteri (2008) the distribution of that range of spins as a function of redshift.
Because accretion is built into the model, we must limit ourselves to extracting the results from mergers and we do so by appealing to the isotropic simulations 
(see Berti \& Volonteri 2008 Figure 4 for details).  We estimate that roughly half of accreting black holes have spins in the required range for them to be FSRQs across cosmic time 
(i.e. negligible dependence on redshift). Hence, we multiply the numbers in column 2 of Table 1 by 0.5. Because we have assumed that black holes with masses equal to or greater than 
$10^{9}$ solar masses are stable with respect to their accretion disk orientation, our black holes are equally likely to end up in prograde or retrograde configurations which means
we must multiply by an additional 0.5. Therefore, our systematically shifted prediction for the number of FSRQs as a function of redshift requires that we multiply the numbers in 
Table 1 column 2 by $\frac{1}{4}$. At this point we attempt to normalize by imposing that our maximum predicted number of FSRQs be equal to the maximum observed number of FSRQs 
(red curve in Figure 3). This requires that we multiply all the values in Table 1 column 2 by $\frac{1}{4}$ and then by 5.56. The results appear in Table 1 column 3. 
By construction then, we see that both our predicted FSRQ curve and the observed FSRQ curve both peak at the same value of 40, our peak shifted to higher redshift as noted (Figure 3).

\begin{figure}
   \centering
   \includegraphics[width=\textwidth, angle=0]{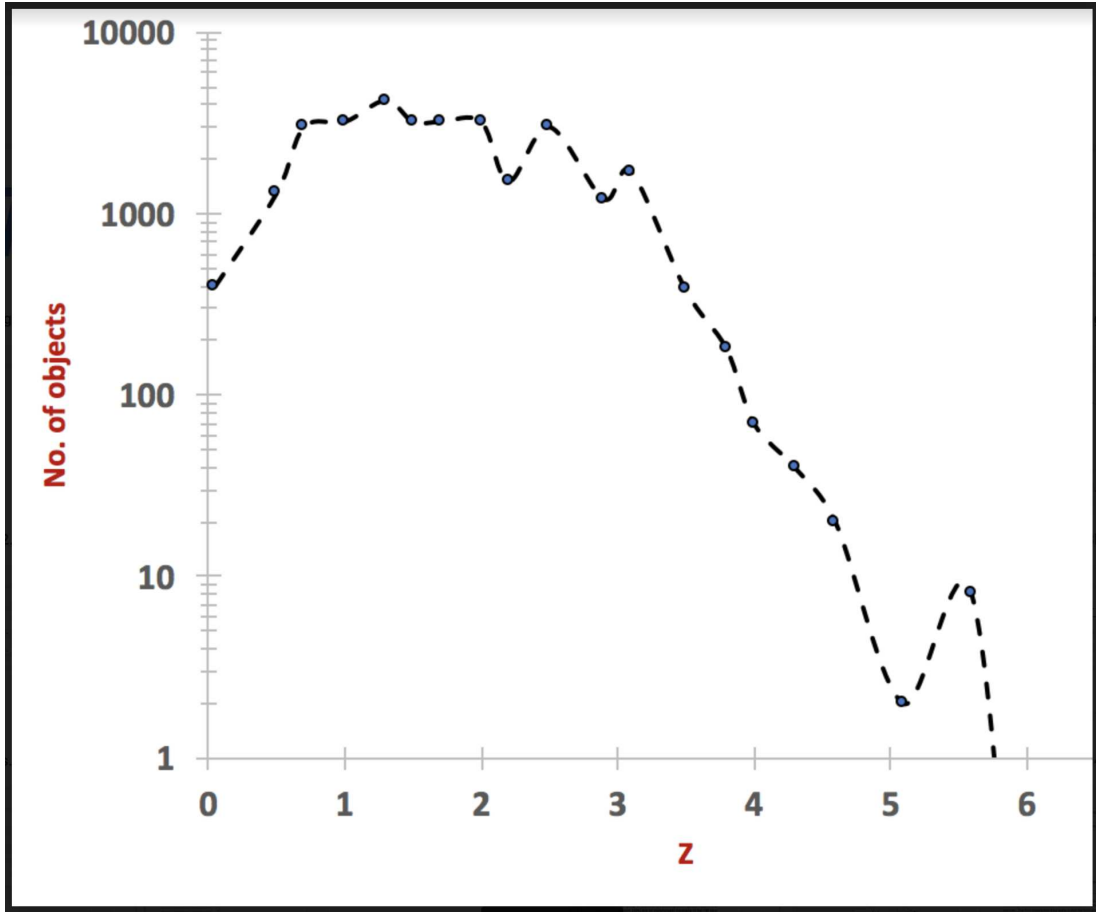}
   \caption{Fermi-detected FSRQs assuming all BHs satisfying the requisite conditions as determined by GD18 including  $dM/dt > 0.01$ produce an FSRQ jet, smoothed version of plot in 
    GD18 (from Shaw et al 2013). }
   \label{Fig1}
   \end{figure}

\begin{figure}
   \centering
   \includegraphics[width=\textwidth, angle=0]{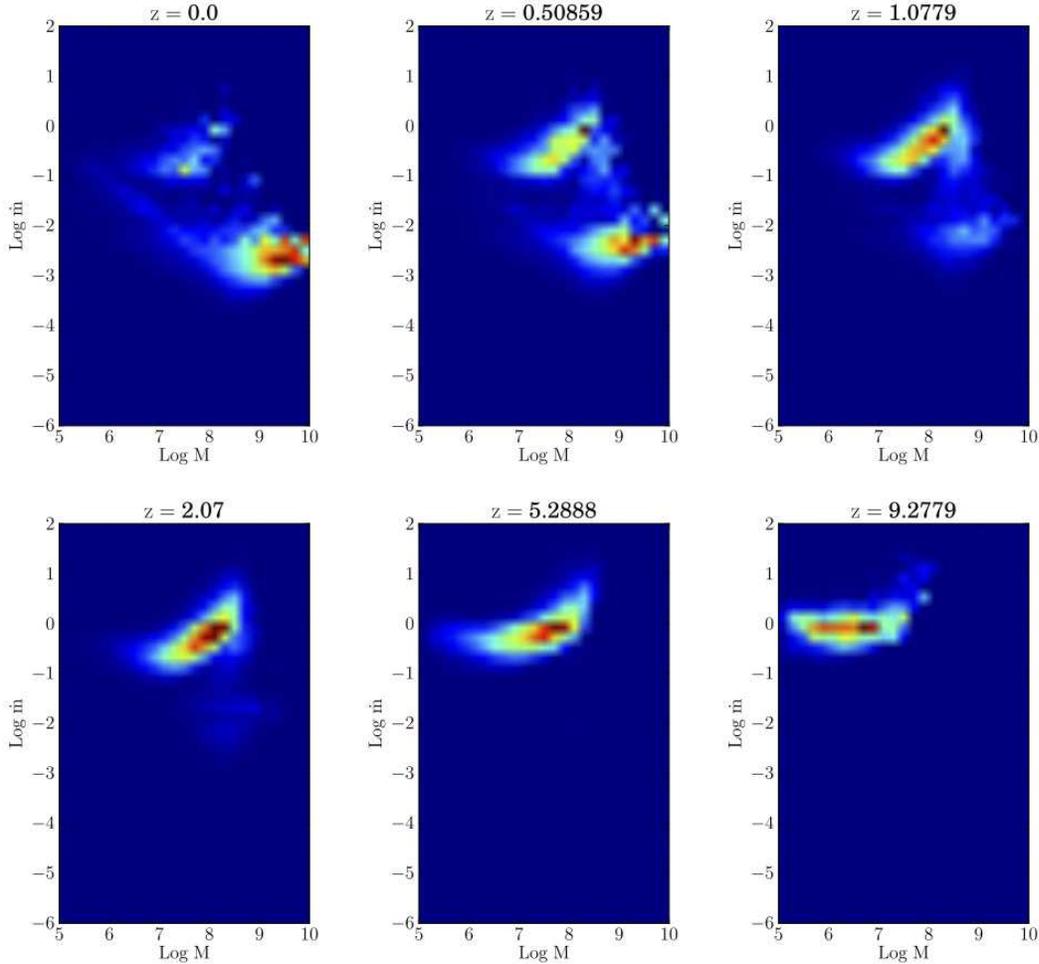}
   \caption{Millennium Simulation prediction for the distribution of accretion as a function of black hole mass for six different redshift values, taken from Figure 4 of Gardner \& Done (2018).
   Colors trace luminosity density as described in GD14 and GD18. The number of accreting black holes with masses greater than or equal to $10^{9}$ solar mass at each of these redshifts 
   is reported in Table 1 column 2, courtesy of C. Done and E. Gardner.}
   \label{Fig2}
   \end{figure}

\begin{figure}
   \centering
   \includegraphics[width=\textwidth, angle=0]{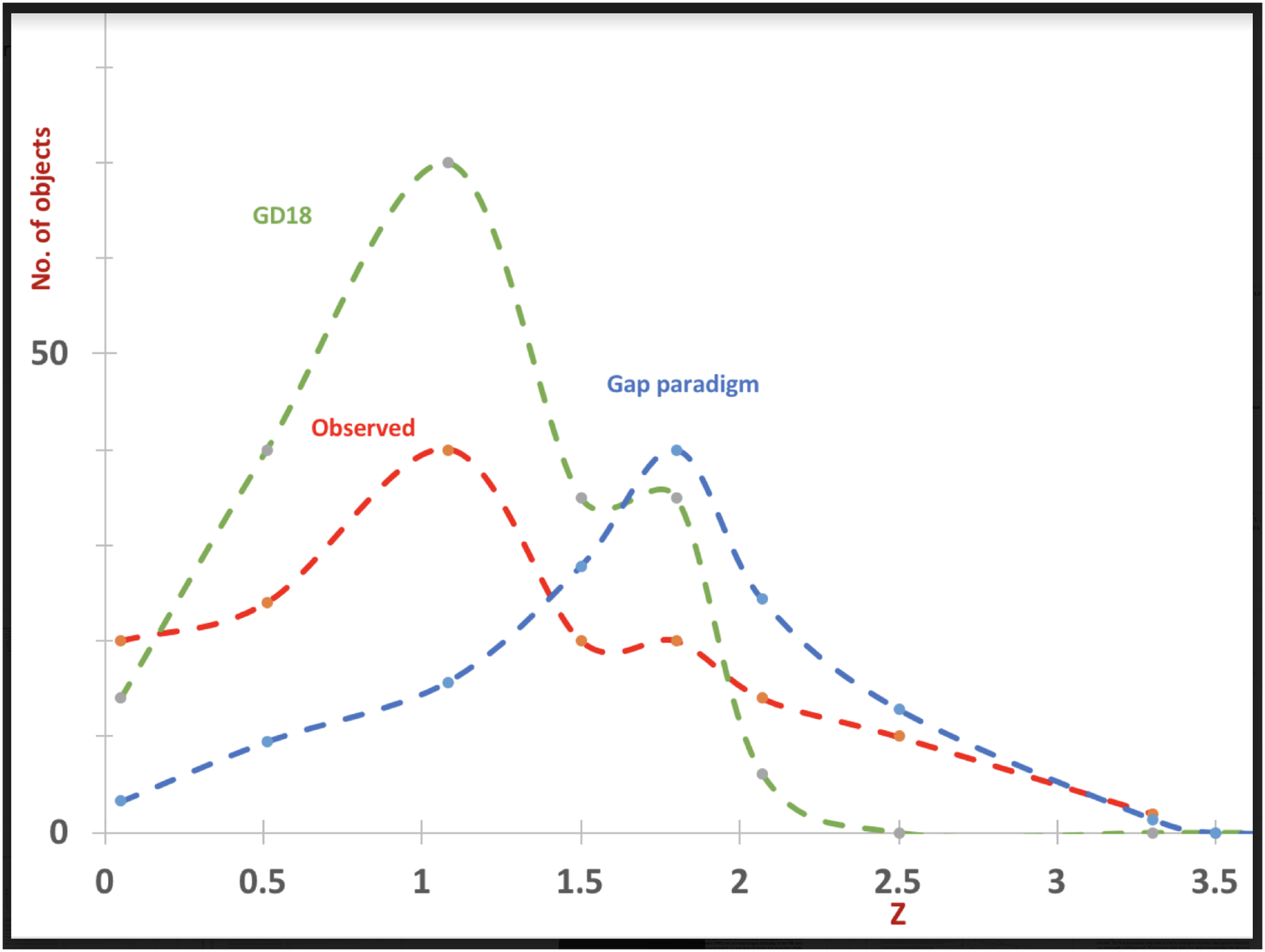}
   \caption{Number of Fermi visible FSRQs in red, predicted number of Fermi visible FSRQs in the gap paradigm in blue, and prediction of FSRQs from GD18 in green as a function of redshift. 
   Functions are smoothened to facilitate comparison.}
   \label{Fig3}
   \end{figure}

Our results in blue are plotted in Figure 3 alongside the Fermi observed numbers in red, and the prediction from Gardner \& Done in green. To summarize, both the green and blue curves
are obtained by constraining the black curve of Figure 1 using different assumptions. For the green curve the constraint is black hole spin above 0.77 while for the blue curve it 
involves half of the population above $10^{9}$ solar masses over a wider range of spins.

\section{BL Lacs}
\label{sect:bllacs}
The gap paradigm for black hole accretion and jet formation connects FRI radio galaxies -  the parent population of BL Lacs - to FRII quasars - the parent population of FSRQs. 
There are two numbers that we must determine in order to produce a version of Figure 3 for predicted number of BL Lacs. The first number determines the fraction of FSRQs that 
evolve into BL Lacs which allows a predicted number for BL Lacs. The second determines the time it takes the FSRQ to evolve into a mature BL Lac which leads to a predicted 
redshift value for BL Lacs. Both numbers will come from a straightforward adherence to the phenomenology illustrated in Figures 4, 5, and 6, which serve to convey the basic 
time dependence in the paradigm (Garofalo, Evans \& Sambruna 2010) which we now describe.\\

If the jet is not powerful enough, accretion remains in a radiatively efficient state and the radio quasar evolves into a radio quiet quasar as the black hole and disk co-rotate.
This is illustrated in Figure 6. At the other end of the spectrum, powerful jets rapidly affect the accretion state which turns the radio quasar into a radio galaxy as seen in 
Figure 4. And in-between these two extremes, we have a jet that more slowly affects the accretion mode compared to Figure 4 but eventually does so unlike Figure 6, as shown in 
Figure 5. Because the timescale for evolution depends on jet power, which in turn depends on black hole mass, what is needed is a prescription that allows one to go from the 
range of accreting black hole masses to a determination of the range of jet powers. Once we know the range of jet powers, we can estimate the fraction of FSRQs that travel along 
paths described in each of Figures 4, 5, and 6. The greater the fraction characterized by Figure 4, the higher the number of predicted BL Lacs. The greater the fraction of objects 
that follow the path described by Figure 6, the lower the number of predicted BL Lacs. For simplicity and to get our bearings, let us assume that accretion versus redshift is scale 
invariant so that all black holes regardless of mass experience the same degree of accretion. In that case, about $\frac{1}{3}$  of the FSRQs would be subject to the evolution 
depicted in Figure 4, $\frac{1}{3}$ to that of Figure 5, and $\frac{1}{3}$ to that of Figure 6, which in turn implies that only $\frac{2}{3}$ of the FSRQs would become BL Lacs. 
We would then multiply the number of FSRQs by $\frac{2}{3}$ in order to obtain a number for the BL Lacs curve. However, our situation does not involve scale invariance which means 
accretion distributes itself differently across the black hole mass scale at different redshifts, and our analysis requires figuring out the fraction of objects following the three 
evolutionary scenarios, which thus is not simply $\frac{1}{3}$ each. In practice, the range of possible paths is not rigidly divided into three classes but spans a continuous space.
This will come into play when we make estimates for the BL Lac redshifts.\\ 

The different redshift values for the BL Lac curve are obtained by recognizing that Figures 4, 5, and 6, imply different timescales for the evolution of FSRQs into BL Lacs for the
subset of FSRQs that in fact evolve into BL Lacs. More precisely, consider for example, the Millennium Simulation results at redshift 2 as seen in the lower left of Figure 2. This
is the redshift at which the largest fraction of the most massive black holes are accreting in radiatively efficient mode which means this is the time when the greatest number of 
most powerful radio quasars are produced. Courtesy of C. Done and E. Gardner, we actually have the precise numbers for a greater variety of redshifts as shown in Table 1 column 2. 
We see that the greatest number of retrograde accreting black holes occurs at redshift of 1.8 and drops off above and below that redshift. Figure 4 matters more during this time 
than at any other in the sense that a greater number of  objects follow this path. Figure 4 also illustrates the presence of an ADAF while the system is still in retrograde mode 
(i.e. with $a = - 0.5$).  But, according to the gap paradigm, BL Lacs are part of the FRI radio galaxy family so the system needs to go through zero spin and then up into the 
prograde spin regime to enter the BL Lac family. ADAF states take longer time to spin up their black holes compared to radiatively efficient mode, possibly by orders of magnitude. 
At the Eddington limit, a maximally spinning black hole in retrograde configuration will be spun down to zero spin in $8 \times 10^{6}$ years (Kim et al 2016 and references therein)
but the quick transition to an ADAF state for Figure 4 objects means this timescale is enhanced by up to a few orders of magnitude even assuming continued accretion. In short, 
the points on the predicted FSRQ versus redshift plot representing the most massive black hole population become points on the BL Lac curve that are most shifted toward lower 
redshift (Figure 7 blue curve).\\ 

By contrast, FSRQs with a range of lower black hole masses also tend to have lower jet powers, and therefore subject to the evolution described either in Figures 5 or 6. 
But to contribute to Figure 7, we must be dealing with objects that do indeed make it to the BL Lac stage which means we are interested only in objects that follow the paths 
of Figures 4 and 5, and not 6. The FSRQs that follow the path of Figure 6, in fact, never become BL Lacs because their jets become suppressed (Neilsen \& Lee 2009; Garofalo, 
Evans \& Sambruna 2010; Ponti et al 2012; Garofalo \& Singh 2016). As an aside, objects that follow paths as in Figure 6 as well as postmerger prograde radiatively efficient 
disks all end up part of the radio quiet quasar/AGN population (Garofalo et al 2016). For the objects described in Figure 5, the transition to the ADAF stage occurs more 
slowly compared to objects described by Figure 4 (because the ADAF stage is reached later).  Hence, the objects belonging to the Figure 5 class evolve more quickly into 
BL Lacs from FSRQs. It might save the reader confusion to point out that objects labeled FRII LERG are not a parent population of BL Lacs. If they were, the timescales between 
objects in Figures 4 and 5 would be reversed. The objects that are not characterized by the largest black holes transition to BL Lacs with smaller fractions compared to those in 
Figure 4 because a higher fraction of them evolve according to Figure 6 and therefore never enter the BL Lac classification. Overall, the connection postulated between FSRQs and 
BL Lacs produces a BL Lac versus redshift curve with the most massive black hole population squeezed up and peaking at lower redshift compared to a less massive black hole 
population represented by a flatter curve at larger redshift.\\    

The basic strategy we have described above can be summarized in the following way. The Millennium Simulation tells us the range of mass for cold mode accreting black holes as 
a function of redshift. For a given range of mass, a weighted contribution must be determined between the three paths of Figures 4-6. For redshifts where the range of accreting 
black holes is smaller, the population will follow paths that involve a weighted average between Figures 5 and 6. For redshifts where the range of accreting black holes is larger, 
the population will follow paths that involve a weighted average between Figures 4 and 5.\\ 

The point by point analysis of each predicted BL Lac number versus redshift is reported in Table 2 and determined as follows. From the first data point for the predicted no. of FSRQs
of 3.22 at redshift of 0.05, we need the fraction that become BL Lacs. In the third column of Table 2 we report a fraction stemming from the distribution of accreting black hole mass 
at each redshift. Because the heaviest black hole population is distributed between redshift 1.5 and 1.8, the FSRQs forming during this time are the ones with the greatest number of 
most powerful jets. Hence, a greater fraction of this population of black holes will make it to the BL Lac phase. But the mass distribution at redshift of 0.05 is not focused on the 
$10^{9}$ solar mass objects. Hence, few of the FSRQs formed at this time evolve into BL Lacs. Our estimate for the fraction of FSRQs forming at redshift of 0.05 that make it to BL Lacs is 0.1.\\   

Hence, we multiply $3.22$ by $0.1$ to get $0.322$ as the predicted number of BL Lacs as shown in Table 2 column 4. The next task is to determine the redshift associated with this number which means
we need the time it takes for these $0.322$ FSRQs to become mature BL Lacs. Because the progenitor FSRQ population is dominated by the less massive of the retrograde black holes, the $0.322$ 
that do become BL Lacs will do so by following paths similar to that described in Figure 5. Because these paths are characterized by radiatively efficient accretion, the characteristic 
timescale for evolution into the prograde regime is near the Eddington value which is in the tens of millions of years. We choose 100 million years. Such timescale adds very little to 
the redshift value which therefore appears roughly the same as for the FSRQs, namely $0.04$. The first data point for the BL Lac curve is therefore 0.322 at redshift $0.04$.\\ 

The second data point for the BL Lac curve originates in a predicted FSRQ number of 9.48. Because accretion clusters onto more massive black holes at this redshift compared to the first
row of Table 2, the fraction of this population that makes it to the BL Lac phase increases. From the data from Gardner \& Done, we increase the fraction to $\frac{1}{3}$ to take into account 
the presence of a greater number of more massive black holes, and thus more massive and more powerful FSRQ jets. Hence, we multiply the original number of predicted FSRQs  which is 
9.48 by $\frac{1}{3}$ to obtain 3.16 for the predicted number of BL Lacs. Because more of the heavy black holes are represented compared to the first row of Table 2, the timescale is a little slower.
Our estimate is 500 million years which results in a redshift value for the objects to emerge as full-fledged BL Lacs of 0.3. The third data point at $15.73$ in Table 2 column 1 
corresponds to a redshift value of 1.08 when it is an FSRQ. Given the larger distribution of black hole mass for this population, more paths of the kind described by Figure 4 appear 
and the timescale is slower. We estimate 1 billion years for the average black hole for the progenitor FSRQs to become BL Lacs. This results in a redshift value of 0.8. The FSRQs forming 
at redshifts in the range 1.5-2.07 are characterized by the most massive black holes and therefore by Figure 4. The timescale for becoming mature BL Lacs is estimated at 6 billion years 
for all three groups. From the original FSRQ redshifts of 1.5, 1.8, and 2.07, their BL Lac counterparts emerge at redshifts 0.4, 0.49, and 0.6, respectively. The remaining rows are 
characterized by smaller black holes which therefore evolve quickly into their BL Lac counterparts, namely from redshifts 2.5, 3.3 to 2.2 and 3.3.\\ 

Our data is shown and compared to the Fermi observed number and GD14 prediction in Figure 7. Notice, however, the interesting fact that the different timescales for evolution has 
changed the order for the redshifts in the last column for the BL Lacs compared to the order in the second column for the FSRQs (i.e. the last column now reads 0.04, 0.3, 0.8, and 0.4). 
This means that our model predicts that some FSRQs forming earlier than others will become mature BL Lacs later. We will come back to this feature in Section 4 to highlight what as 
far as we know is an unidentified feature of the observations that we will motivate physically.

\begin{figure}
   \centering
   \includegraphics[width=\textwidth, angle=0]{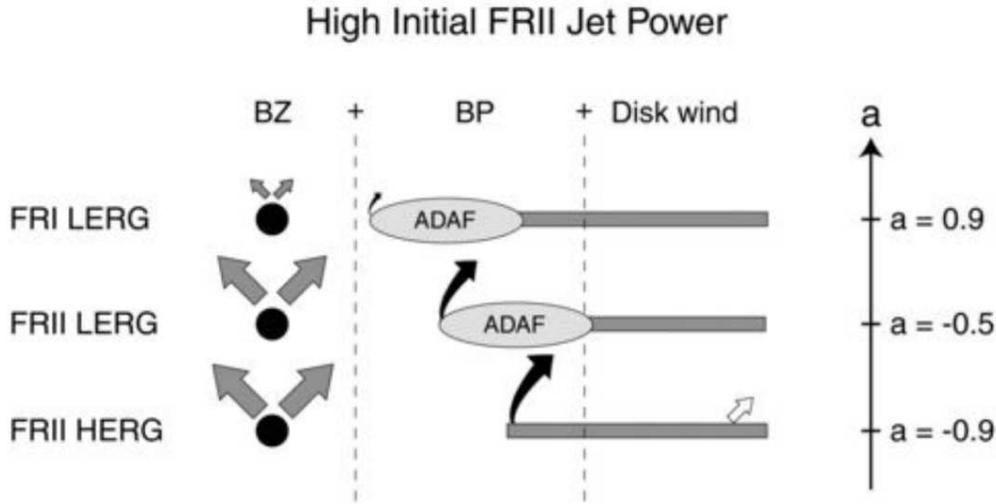}
   \caption{Time evolution of an initially retrograde accreting black hole with powerful jets (from Garofalo et al 2010; where LERG is “low-excitation radio galaxies” and 
   HERG is “high-excitation radio galaxies”). The radiative efficiency of the initially radiatively efficient thin disk (lower panel) evolves quickly into an advection-dominated disk, 
   which has a radiative efficiency that is at least two orders of magnitude smaller than the Eddington value. The jet is the result of a combination of the Blandford-Znajek process
   (BZ - Blandford \& Znajek 1977) and the Blandford-Payne process (BP - Blandford \& Payne 1982) which explains the labels in the first and second columns. In ADAF states the radiative
   disk wind is quenched.}
   \label{Fig4}
   \end{figure}

\begin{figure}
   \centering
   \includegraphics[width=\textwidth, angle=0]{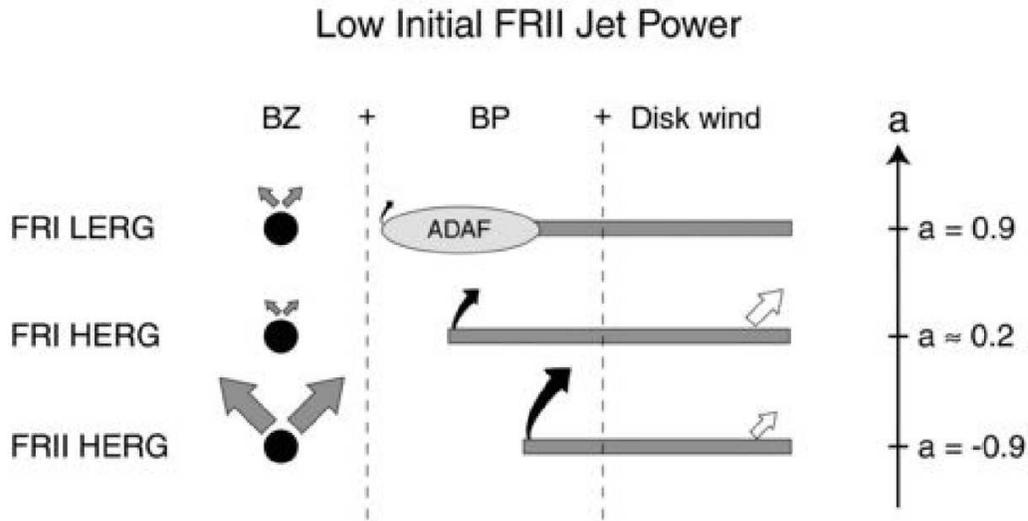}
   \caption{Time evolution of an initially retrograde accreting black hole with less powerful jets (from Garofalo et al 2010; Details same as Figure 4.). The radiative efficiency of 
   the initially radiatively efficient thin disk (lower two panels) evolves less quickly (compared to the object in Figure 4) into an advection-dominated disk.}
   \label{Fig5}
   \end{figure}

\begin{figure}
   \centering
   \includegraphics[width=\textwidth, angle=0]{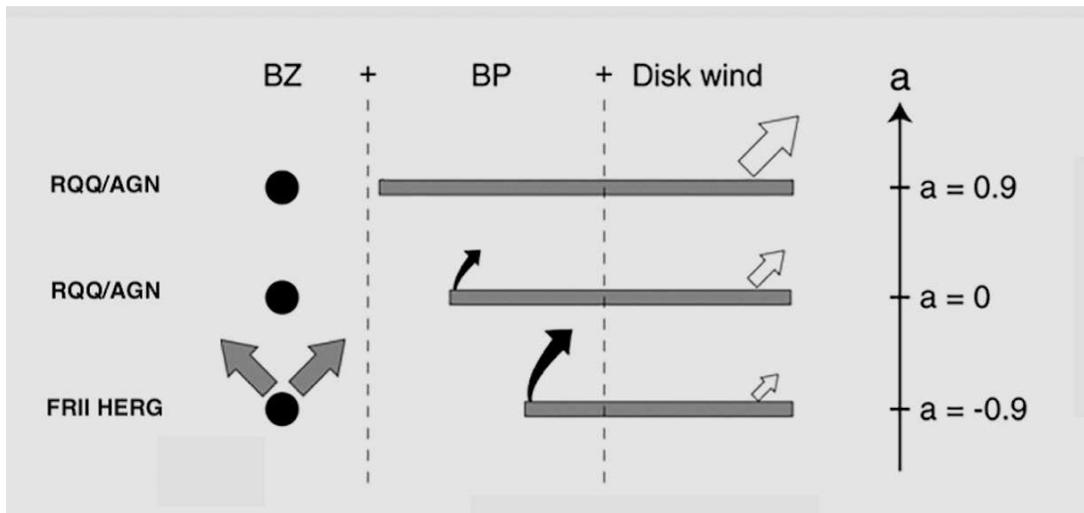}
   \caption{When the FRII quasar jet is not sufficient to alter the accretion mode, the system evolves into a prograde regime while remaining in a radiatively efficient state. 
    As a result of this continued radiative efficiency of the disk, the disk wind suppresses the jet and a radio quiet quasar (RQQ)/AGN emerges.}
   \label{Fig6}
   \end{figure}

\begin{figure}
   \centering
   \includegraphics[width=\textwidth, angle=0]{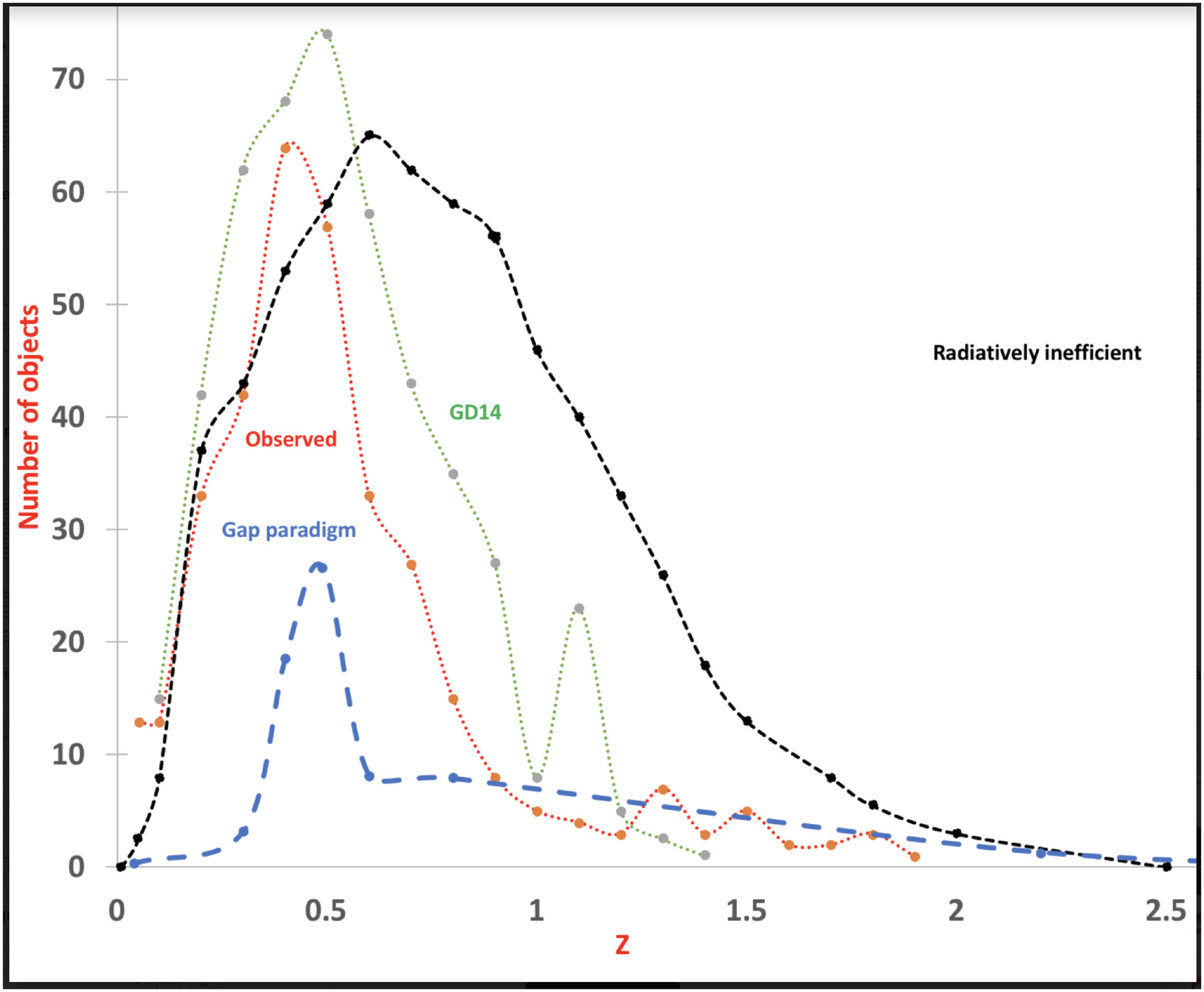}
   \caption{Fermi number of observed BL Lac versus redshift distribution in red, GD14 predicted number of BL Lacs versus redshift distribution in green (labeled GD14), gap paradigm 
    number of BL Lacs predicted versus redshift distribution in blue, and the radiatively inefficient family of objects identified by Gardner \& Done (2014) divided by $100$ in black 
    (i.e. the actual number is two orders of magnitude larger).}
   \label{Fig7}
   \end{figure}

\section{Discussion}
\label{sect:discussion}
Armed with our new redshift distributions shown in Figures 3 and 7, we now address the issues highlighted in the abstract and Introduction.

\subsection{Why the numbers of BL Lacs and FSRQs are of the same order of magnitude ?}
As pointed out by GD14 and GD18 there are about 500 BL Lacs in the Second Fermi Large Area Telescope catalog and about 300 FSRQs in the first Fermi catalogue (Abdo et al. 2010; 
Ackermann et al. 2011). If the type of accretion were solely responsible for the nature of the jet producing object (either FSRQ or BL Lac), the numbers determined from Gardner \& Done 
prior to spin constraints would differ by an order of magnitude with BL Lacs outnumbering FSRQs. As a result, Gardner \& Done further constrain FSRQs and BL Lacs by needed but ad-hoc 
assumptions about black hole spin with dimensionless values $a > 0.77$ and $a > 0.8$ for the two families, respectively, a move that is actually at odds with general relativistic 
magnetohydrodynamic simulations (e.g. Tchekhovskoy et al. 2010; Tchekhovskoy et al. 2011, McKinney et al. 2012) because such simulations generally predict powerful jets only when 
the spin is quite high (i.e. $a > 0.9$ or even much closer to 1).  This lower spin range is necessary since the idea of very high spins in these populations would not restrict their 
numbers sufficiently to match Fermi observations. We have shown that the assumption of retrograde accretion for FSRQs produces a subset of the $dM/dt > 0.01 (dM/dt)_{Edd}$ that does 
a decent job of matching with observations. It is fair to say, however, that although we claim the constraints on spin adopted by GD18 for the FSRQs are not sufficiently motivated, 
their assumption does generate a predicted number of FSRQs that is as decent a match with observations as anything we have done, if not better. Hence, the different prescriptions 
for FSRQ numbers between the two paradigms is not by itself a sufficient reason to prefer our framework, quite the contrary. In contrast to the additional assumptions required to 
lower the large number of predicted BL Lacs in GD14, no assumptions are inserted in our framework at this stage. To determine the BL Lac distribution, in fact, our job is to quantitatively 
determine the model prediction from its phenomenology. Our model, of course, predicts that BL Lacs are fewer than FSRQs but observationally the FSRQs are more difficult to observe 
because predominantly at higher redshift. Again, the order-of-magnitude equality predicted from our framework for the two classes of AGNs is insufficient to compel a preference for 
our description. In fact, our BL Lac prediction comes in a little low compared to GD14 (Figure 7). Instead, as we describe in the next sections, it is the confluence of explanation in 
our framework’s ability to resolve issues 2 and 3 below, that we think produces merit.

\subsection{A higher redshift BL Lac versus redshift tail}
In this section our focus is on a difference between the two observed distributions as a function of redshift that as far as we know has not been singled out. Whereas the observed 
FSRQ versus redshift distribution (red curve in Figure 3) dies down at both high and low redshift in the sense that it could be described as a truncated normal distribution, 
the observed BL Lac versus redshift distribution (red curve in Figure 7) does not appear to similarly truncate at higher redshift but retains a flat tail. Our claim is that such a 
feature is the signature of a connection between the two populations of active galaxies because we find that it emerges naturally from the evolutionary pictures described in 
Figures 4 through 6. In order to appreciate its origin, we need to take a closer look at the data for our predicted BL Lac curve.\\ 

The main effect of this connection between FSRQs and BL Lacs is to produce a peak that moves to lower redshift but with a tail lingering at higher redshift. The peak of the BL Lac 
curve is generated from the values of the FSRQ curve associated with the most massive black holes. Such objects are the slowest to evolve because their powerful jets rapidly alter 
the timescale of evolution due to ADAF accretion so they emerge as mature BL Lacs at lower redshift. Notice, as anticipated in Section 3, how the fourth row in Table 2 starts with 
a higher redshift compared with the third row (when it is associated with FSRQs), but ends up with a lower redshift compared to the third row (when it describes BL Lacs). The simple 
evolutionary picture described in Figures 4 through 6, therefore, has the basic effect of producing a squeezing and a stretching of the FSRQ curve into a BL Lac curve. The squeezing 
of the peak of the BL Lac curve comes from the population with the heaviest black holes. Such objects evolve more slowly than others so they emerge as BL Lacs at lower redshift. But
they are also characterized by the greatest numbers because a greater fraction of the original FSRQs become BL Lacs. Hence, highest vertical numbers but lowest redshift numbers, thus 
a squeezing effect. The objects with smaller black holes, instead, evolve quickly so they become mature BL Lacs on shorter timescales (In Table 2 we see this effect dominate for data 
in rows 1 and 8, where the redshifts for both BL Lacs and FSRQs are similar). This produces a stretching in the horizontal direction. Such objects also have a lower number of BL Lacs 
as a result of the fact that an increasing fraction of this group never makes it into the BL Lac family. Hence, this produces both a lowering in the vertical direction and a horizontal 
stretching or tail associated with the peak of the BL Lac versus redshift function. It is also important to point out that although the numbers are small, the tail in the BL Lac curve 
versus redshift is more significant because it occurs at lower redshift compared to the FSRQ distribution. This is due to the fact that the span of time for a lower redshift range 
corresponds to a greater time, enhancing the physical significance of the data. It is also true that the uncertainties and small number of available data points make it so that a more 
detailed analysis would produce varying degrees of squeezing and stretching of the BL Lac versus redshift curve. The focus, thus, should be on the existence of the general effect and 
not on the exact curve. It is worth noting that a BL Lac tail is visible in the 2LAC data of $Figure 12$ in Ackermann et al. 2011 (see also Figure 2 of Bauer et al. 2009). Notice, 
also, that neither the black curve nor what you get from it by imposing spin constraints - the green curve - in Figure 7 show signs of a lower redshift/higher redshift bunching/tail 
behavior. The point being that radiative efficiency and spin constraints do not lead to squeezing and tail-like behavior. Whereas the redshift distribution discussed in this work comes 
from our model prescription, which we have compared to observations, measured redshifts for BL Lacs are difficult to obtain because of the lack of emission lines (e.g., Paiano et al. 2017). 
Despite this, there is evidence of a good match in redshifts between emission lines and objects to which emission line measurements of redshift cannot be attributed, and of small error 
estimates for our BL Lacs redshifts (Mao \& Urry 2017).

\begin{table}
\bc
\begin{minipage}[]{100mm}
\caption[]{The number of predicted BL Lacs is obtained by multiplying the number of predicted FSRQs by the fraction of FSRQs that become BL Lacs. The redshift at which they become
mature BL Lacs depends on the mass distribution or weight of the black holes at each FSRQ redshift.}\end{minipage}
\setlength{\tabcolsep}{1pt}
\small
 \begin{tabular}{c|c|c|c|c|c}
  \hline\noalign{\smallskip}

Predicted No. of FSRQs & Redshift for the FSRQs & Fraction of FSRQs that  &  Predicted No. of  & Timescale for evolution into   & Redshift for the \\
(from Table 1)         & (from Table 1)         &  become BL Lacs         &    BL Lacs         &   a BL Lac (in years)          & BL Lacs\\
  \hline\noalign{\smallskip}
3.22  & 0.05 & 0.1  & 0.322 & 100 million & 0.04\\
9.48  & 0.51 & 0.33 & 3.16  & 500 million & 0.3\\
15.73 & 1.08 & 0.5  & 7.87  & 1 billion & 0.8\\
27.86 & 1.50 & 0.66 & 18.57 & 6 billion & 0.4\\
40    & 1.8  & 0.66 & 26.65 & 6 billion & 0.49\\
24.35 & 2.07 & 0.33 & 8.12  & 6 billion & 0.6\\
12.87 & 2.5  & 0.1  & 1.29  & 380 million & 2.2\\
1.39  & 3.3  & 0.1  & 0.139 & tens of millions & 3.3\\
0     & 3.5  & 0    & 0     & tens of millions & 3.5\\

   \noalign{\smallskip}\hline
\end{tabular}
\ec
\end{table}

\subsection{No random injection of high black hole spin}
As GD14 point out, the assumption that all radiatively inefficient accreting black holes produce a BL Lac, overestimates the observed number by three orders of magnitude. Clearly, 
further constraints are needed to better match the observations. GD14 assume black hole spin is a natural candidate and both arbitrarily and probably insufficiently (from the point 
of view of jet physics), choose a value of 0.8 as a threshold value for producing a BL Lac . As Figure 7 shows, this produces a better match with Fermi observations. However, this 
is accomplished by assuming that chaotic accretion operates to produce low spins for most of the AGN population at lower redshift while mergers are responsible for the high spins 
needed for BL Lacs. But the evidence is actually against mergers as the recent triggering mechanism in FRI radio galaxies (Heckman et al. 1986; Baum, Heckman \& van Breugel 1992; 
Chiaberge et al. 1999; Hardcastle, Evans \& Croston 2007; Baldi \& Capetti 2008; Emonts et al. 2010; Ivison et al. 2012), of which BL Lacs are a subset. In addition, as already
pointed out, the idea that black hole spin values of $a > 0.8$ is enough for the powerful BL Lac jets is not supported by either analytic or numerical work which point, instead, 
to very high spins for the most powerful jets with $a > 0.9999$ (Tchekhovskoy et al. 2010), with  $a > 0.99$ (Tchekhovskoy et al. 2011) or even in the flooded magnetospheres 
with strongest magnetic fields requiring $a > 0.9$ (McKinney et al. 2012). The picture of Gardner \& Done must also implicitly view evidence of high black hole spins in tens of 
radio quiet quasars and AGN at low redshift as faulty (e.g. Brenneman 2013 and references therein). In our phenomenological framework, on the other hand, both the high (prograde) 
spin values as well as the origin of hot mode accretion, emerge from the same place, namely progenitor FSRQs triggered in the aftermath of a merger that evolve via accretion. 
Incidentally, because most of the accreting black holes over cosmic time do not meet the mass requirement and therefore are triggered by mergers into radio quiet-like mode, we 
also have an explanation for the origin of LINERs and why they dominate at low redshift over other classes of AGN (see Garofalo et al 2016 for a deeper discussion of this).

\section{Conclusions}
\label{sect:conclusion}
We have used the analysis of GD18 in determining the number of accreting black holes satisfying $dM/dt > 0.01 (dM/dt)_{Edd}$ as the starting point  for prediction of the observed 
numbers of FSRQs based on the Millennium Simulation and from it appeal to the time evolution prescribed in the gap paradigm to produce a BL Lac versus redshift curve. While neither 
our FSRQs versus redshift nor our BL Lacs versus redshift curves are an improvement compared to the predictions of GD14 and GD18, we argue that ours qualitatively reproduces a 
number of observational features invoking the fewest ad-hoc assumptions in a unified way. We have singled out three requirements that appear necessary in order to explain the 
observations, namely, the equal order of magnitude in number of observed FSRQs and BL Lacs, a bunching up at low redshift and a tail at higher redshift for the BL Lacs, and 
a need for large black hole spins at significantly separated cosmic times. We have shown how all three of these features emerge from a single idea. As deeper surveys become 
available, support for the model in terms of population numbers and redshift difference in FSRQs and BL Lac can be checked (e.g. in FL8Y compared to surveys with fewer blazars 
such as in Fermi LAT 3FGL). In closing, we also emphasize an implication of the model which is that the mass of a black hole that is originally high retrograde and ends up in 
the high prograde regime, increases by a factor of about 2.5. The model, therefore, also predicts that BL Lacs will tend to not only appear at lower average redshift, but also 
with average black hole masses that are larger than those of FSRQs.

\begin{acknowledgements}
We thank C. Done for providing the data to the Millennium Simulation. CBS was supported by the I-Core centre of excellence of the CHE-ISF.
\end{acknowledgements}

\label{lastpage}

\end{document}